\documentclass[a4paper,11pt]{article}
\usepackage{pos}
\usepackage{slashed}

\usepackage{graphicx}  % needed for figures
\usepackage{dcolumn}   % needed for some tables
\usepackage{bm}        % for math
\usepackage{physics}
\usepackage{tikz}
\usepackage{mathtools}
\usepackage{amsmath} %maths

\newcommand{\FTEE}{FTE$^2$ }
\newcommand{\FTEEformat}{FTE$^2$}

\newcommand{\QQbar}{{Q\bar Q}}

\newcommand{\lp}{\left}
\newcommand{\rp}{\right}

\FullConference{The 40th International Symposium on Lattice Field Theory (Lattice 2023)\\
July 31st - August 4th, 2023\\
Fermi National Accelerator Laboratory\\}
\title{Internal color contributions to flux tube entanglement entropy}
\author*[a]{Rocco Amorosso}
\author[a]{Sergey Syritsyn}
\author[b,c]{Raju Venugopalan}

\affiliation[a]{Department of Physics and Astronomy, Stony Brook University, 
  Stony Brook, New York 11794, USA}
\affiliation[b]{Physics Department, Brookhaven National Laboratory, Upton, NY 11973, USA}
\affiliation[c]{CFNS, Department of Physics and Astronomy, Stony Brook University, Stony Brook, NY 11794, USA}
\date{\today}%

\emailAdd{rocco.amorosso@stonybrook.edu}
\emailAdd{sergey.syritsyn@stonybrook.edu}
\emailAdd{raju.venugopalan@stonybrook.edu}

\abstract{
In recent work~\cite{Amorosso:2024leg}, we introduced and computed entanglement entropy of the color flux tube (FTE$^2$) between a heavy quark-antiquark pair in (2+1)D Yang-Mills theory. Our numerical results suggest that FTE$^2$ can be partitioned into a component corresponding to transverse vibrations of the flux tube and an internal color entropy. 
Further, motivated by analytical (1+1)D calculations, and  SU(2) (2+1)D Yang-Mills numerical results, we argued that the internal entropy 
takes the form $\langle F\rangle\log(N_c)$, with $\langle F\rangle$ the number of times, on average, that the flux tube crossed a boundary between region $V$ and its complement. We extend here our FTE$^2$ study to consider different geometries of region $V$, varying the number of boundary crossings, and number of colors.
Our preliminary results support the conjectured form of the internal entropy, albeit with noteworthy subtleties relating to the partial/full intersections of the flux tube with the region $V$.
}

\FullConference{The 41st International Symposium on Lattice Field Theory (LATTICE2024)\\
28 July - 3 August 2024\\
Liverpool, UK\\}

\begin{document}
\maketitle

%%%%%%%%%%%%%%%%%%%%%%%%%%%%%%%%%%%%%%%%%%%%%%%%%%%%%%%%%%%%%%%%%%%%%%%%%%%%%%
\section{Introduction}
Quantum entanglement is an intrinsic feature of quantum systems, reflecting non-classical correlations between subsystems.
The entanglement entropy is computed from the reduced density matrix
\begin{equation}
\label{eqn:rho_reduced}
\hat{\rho}_V=\Tr_{\bar{V}}\hat{\rho}\,,
\end{equation}
where $\hat{\rho}$ is the full density matrix of the system and the partial trace $\Tr_{\bar{V}}$ is taken over fields in region $\bar{V}$.
The reduced density matrix allows us to define the von Neumann entanglement entropy, 
\begin{equation}
   S^\text{EE}=-\Tr \lp(\hat{\rho}_V \log \hat{\rho}_V\rp).
\end{equation}
Entropy and quantum information have broad explanatory power, allowing us to explore complex systems without explicitly considering intricate microscopic dynamics.
Recently,  interest in quantum entanglement has been growing in nuclear physics, with applications spanning the structure of hadrons~\cite{Kharzeev:2017qzs,Beane:2018oxh}, the dynamics of hadronization~\cite{Kharzeev:1996sq,Berges:2017zws,Gong:2021bcp,Barata:2023jgd} and thermalization~\cite{Berges:2020fwq,Davoudi:2024osg}, helicity of partons \cite{Bhattacharya:2024sno}, and heavy quarkonium \cite{Kharzeev:2014pha}.
Understanding confinement in the language of quantum information is not without difficulty, since it is 
 well established that calculating the entanglement entropy in lattice gauge theory is highly nontrivial.
Firstly, explicitly calculating the density matrix or the von Neumann entanglement entropy on the
lattice is not feasible, so instead one typically uses the replica trick to study the Renyi entropy of order $q$:
\begin{equation}
S^{(q)}=\frac{1}{1-q}\log\lp(\Tr\hat{\rho_V}^q\rp) .
\end{equation}
In the limit $q\to1$, one recovers the von Neumann entanglement entropy.

More importantly, defining entanglement entropy in a gauge-invariant way is not straightforward.
Due to Gauss's law, the Hilbert space of gauge invariant states cannot be factorized into gauge-invariant subspaces, making the reduced density matrix defined in Eq.~(\ref{eqn:rho_reduced}) ambiguous.
The entanglement entropy in quantum field theories is also typically UV-divergent, scaling with the area for $D>2$ \cite{Bombelli:1986rw,Srednicki:1993im}, so one typically has to study finite associated quantities rather than the Renyi entropy itself. 
Despite these well-known difficulties, entanglement entropy has been studied on the lattice, with previous lattice studies addressing entropic C-functions and vacuum entanglement \cite{Bulgarelli:2024onj,Buividovich:2008gq,Buividovich:2008kq,Itou:2015cyu,Rabenstein:2018bri,Jokela:2023rba}.
In this work, we explore an alternative method for calculating the entanglement entropy in gauge theories, dubbed flux tube entanglement entropy (\FTEEformat) \cite{Amorosso:2023fzt,Amorosso:2024glf,Amorosso:2024leg}, defined as the excess entanglement
entropy relative to the vacuum that can be attributed to the color flux tube stretching between a static quark-antiquark
pair in pure gauge Yang-Mills theory.
In Ref.~\cite{Amorosso:2024leg}, we showed that \FTEE is finite, gauge-invariant, and free of ambiguities in calculating the entanglement entropy related to the the non-factorizability of the Hilbert space.
In this work, we will further explore \FTEEformat, investigating specifically the entropic dynamics of the colorful degrees of freedom inside the color flux tube.

%%%%%%%%%%%%%%%%%%%%%%%%%%%%%%%%%%%%%%%%%%%%%%%%%%%%%%%%%%%%%%%%%%%%%%%%%%%%%%
\section{Flux Tube Entanglement Entropy (\FTEEformat)}
In this section we give a brief overview of flux tube entanglement entropy (\FTEEformat) and its conjectured decomposition into internal and vibrational components.
For a more detailed review of \FTEE in (2+1)D and (1+1)D, we refer the readers to Refs.~\cite{Amorosso:2024leg,Amorosso:2024glf}.
We begin by outlining the replica method and the explicit formulation of \FTEE on the lattice.
We then discuss the vibrational and internal components of \FTEEformat, particularly focusing on the (1+1)D results that motivate the internal entropy and its peculiar features in (2+1)D.

\subsection{Defining \FTEE}

\begin{figure}[ht!]
  \centering
  \includegraphics[width=.4\textwidth]{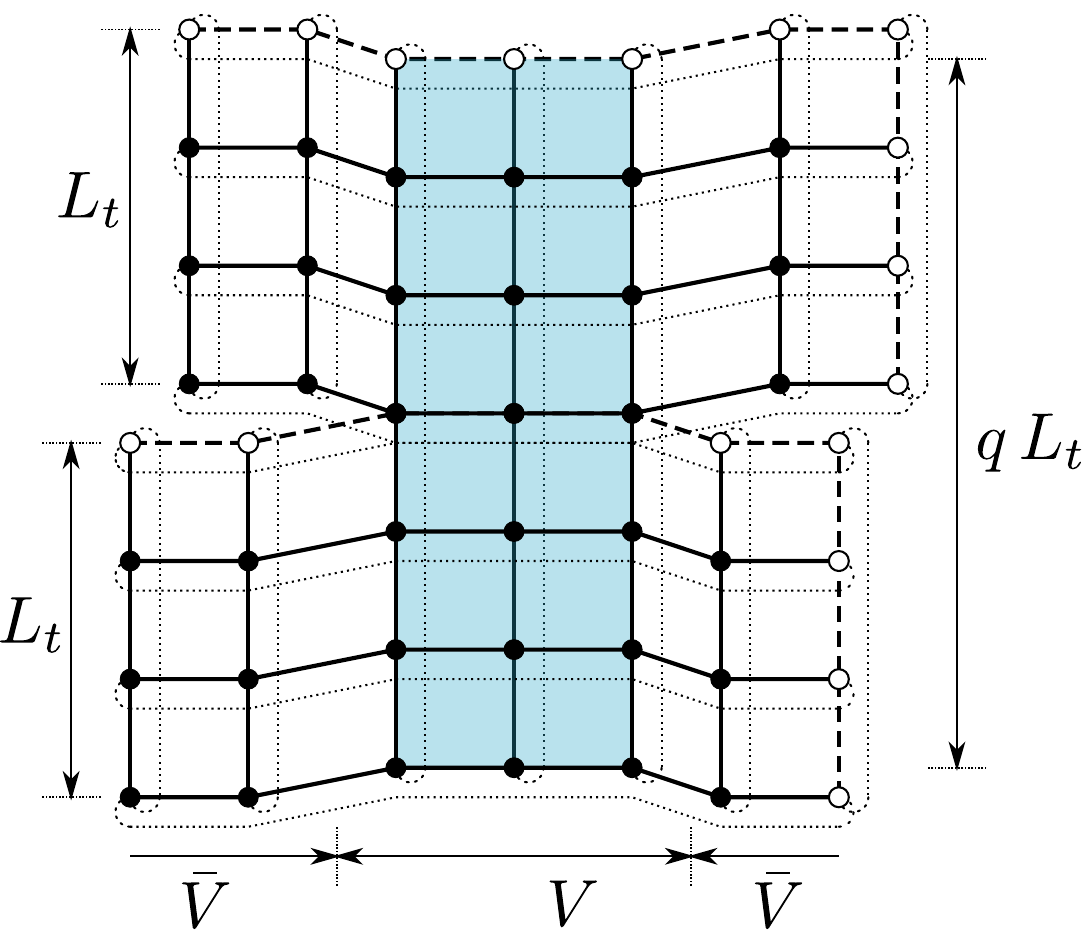}
  \caption{\label{fig:pants}
Depiction of the replica geometry of a $L_x\times L_\tau=6\times3$ lattice with $q=2$ replicas. 
The lattice has periodicity $L_x$ in the spatial direction. 
The lattice is $L_\tau$-periodic in region $\bar{V}$ and $qL_\tau$-periodic in region $V$. 
Open points on edges are wrap-around images of the sites they are 
connected to by dashed lines.
}
\end{figure}

Flux tube entanglement entropy \cite{Amorosso:2023fzt,Amorosso:2024glf,Amorosso:2024leg} is defined as the excess entanglement
entropy relative to the vacuum that can be attributed to the color flux tube.
The entanglement entropy in field theories usually takes the form \cite{Buividovich:2008kq,Ryu:2006bv,Ryu:2006ef,Nishioka:2006gr,Klebanov:2007ws}
\begin{equation}
\label{eq:entropyUVfinite}
S=S_{UV}+S_f\,,
\end{equation}
where $S$ represents the  entropy, $S_f$  the finite part of the entropy, and $S_{UV}$  the UV-divergent component of the entropy.
To extract information about $S_f$, one typically calculates the entropic C-function, taking the derivative of the entropy while keeping the area of the entangling surface constant, effectively canceling contributions from $S_{UV}$.
When studying color flux tubes, one can instead explicitly subtract out the UV-divergent piece.
This is done by subtracting the entanglement entropy of the vacuum from the entanglement entropy of a $Q\bar{Q}$ pair and its confining flux tube.
\FTEE then takes form \cite{Amorosso:2024leg}
\begin{equation}
\label{eqn:RenyiDiff}
\tilde{S}^{(q)}_{\vert Q \bar{Q}} \equiv S^{(q)}_{\vert Q \bar{Q}}-S^{(q)} .
\end{equation}
\FTEE can be calculated on the lattice with Polyakov lines and the replica method.
For computing Renyi entanglement entropy on the lattice, we follow the procedure of
Ref.~\cite{Itou:2015cyu,Rabenstein:2018bri} to generate 
$(\hat{\rho}_V)^q$.
As laid out in Ref.~\cite{Calabrese:2004eu}, $q$ independent replicas of the lattice, each with $L_\tau=\beta=T^{-1}$, are stacked in the temporal direction.
In region $\bar{V}$, links are $L_\tau$ periodic with 
$U_\mu(\tau=0)^{(r)}=U_\mu(\tau=\beta)^{(r)}$ for each replica $r$, implementing the partial trace $\Tr_{\bar{V}}\hat{\rho}$.
In region $V$, link variables are identified 
for consecutive replicas with $U_\mu(\tau=0)^{(r+1)}=U_\mu(\tau=\beta)^{(r)}$ 
and are $qL_\tau$ periodic with $U_\mu(\tau=0)^{(1)}=U_\mu(\tau=\beta)^{(q)}$, corresponding to the product of reduced density matrices and their overall trace of $\rho_V^q$.
This geometry is illustrated in Fig.~\ref{fig:pants} in (1+1)D with $q=2$.

%-----------------------------------------------------------------------------

We place quark and antiquark in $\bar{V}$,
with their color degrees of freedom traced according to Eq.~(\ref{eqn:rho_reduced}),
resulting in $2q$ Polyakov loops of length $L_t$.
Therefore, \FTEE can be calculated directly from the ratio of Polyakov loop correlators as \cite{Amorosso:2024leg,Amorosso:2023fzt}
\begin{align}
\label{eqn:RenyiDiffPolyakovprequel}
\tilde{S}^{(q)}_{\vert Q_{\vec x} \bar{Q}_{\vec y}} = S^{(q)}_{\vert Q_{\vec x} \bar{Q}_{\vec y}} -S^{(q)} 
  &= -\frac{1}{q-1} \left(\log
  \frac{Z^{(q)}_{\vert Q_{\vec x} \bar{Q}_{\vec y}}}{Z^{q}_{\vert Q_{\vec x} \bar{Q}_{\vec y}}}-\log
  \frac{Z^{(q)}}{Z^{q}}\right)
\\
\label{eqn:RenyiDiffPolyakov}
  &= -\frac{1}{q-1} \log
  \frac{\langle \prod\limits_{r=1}^{q}\Tr P_{\vec x}^{(r)}\Tr P_{\vec{y}}^{(r)\dag}\rangle}
       {\big[\langle\Tr P_{\vec x}\Tr P_{\vec{y}}^\dag\rangle\big]^q}
\,,
\end{align}
where $\Tr P_{\vec{x}}^{(r)}$ is the Polyakov loop at location $\vec{x}$ in replica $r$,
and $\Tr P_{\vec{x}}$ is the Polyakov loop on the one-replica lattice.

\subsection{Internal color contributions to FTE$^2$
  \label{sec:models}}
In Ref.~\cite{Amorosso:2024leg} it is conjectured that FTE$^2$ can be partitioned as 
\begin{equation}
\label{eq:vib+int}
\tilde{S}^{(q)}_{\vert Q \bar{Q}} = S_\text{internal} + S_\text{vibrational}\,,
\end{equation}
where $S_\text{internal}$ is the entanglement entropy associated with internal color degrees of freedom, while $S_\text{vibrational}$ is the entanglement entropy associated with the orientation and transverse fluctuations of the flux tube.

To study the vibrational entanglement entropy in Eq.~(\ref{eq:vib+int}), one could model the color flux tube as a quantum vibrating string.
In Ref.~\cite{Amorosso:2024leg}, we examined the predictions of a simple model, the Gaussian approximation for small transverse fluctuations of a thin string \cite{Luscher:1980iy}.
By studying this system, one finds the logarithmic behavior 
$\propto\frac{1}{4}\log(L)$ for the vibrational Rényi entropy with $q=2$ for a flux tube of length $L$.
This behavior was observed on the lattice in Ref.~\cite{Amorosso:2024leg}.

The vibrational entropy, however, for the $Q\bar{Q}$ separations studied in Ref.~\cite{Amorosso:2024leg} (up to $L\sqrt{\sigma}\sim0.8$), is of much smaller magnitude than the internal entropy.
The internal entanglement entropy is the dominant source of entanglement entropy that can be attributed to the flux tube, making up roughly $\sim90-95\%$ of FTE$^2$.
The internal entropy  reflects the possible color states of the flux tube and must be independent of the shape of the flux tube arising from its vibrations.

In Ref.~\cite{Amorosso:2024glf}, it was shown that FTE$^2$ in (1+1)-dimensional
Yang-Mills theory, where a flux tube does not have transverse degrees of freedom, takes the form
\begin{equation}
\label{eqn:1+1analytical}
\big(\tilde{S}^{(q)}_{\vert Q \bar{Q}}\big)_{\rm (1+1) YM}= F\cdot\log (N_c)\,,
\end{equation}
 where $F$ is the number of boundaries between $V$ and $\bar V$ crossed by the flux tube.

\begin{figure}[t!]
\centering
\includegraphics[width=.45\textwidth]{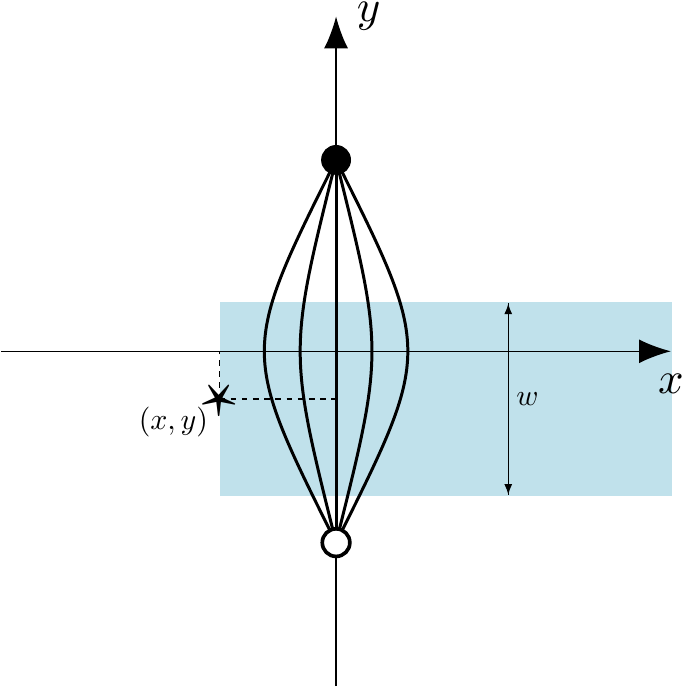}
\hfill
\includegraphics[width=.45\textwidth]{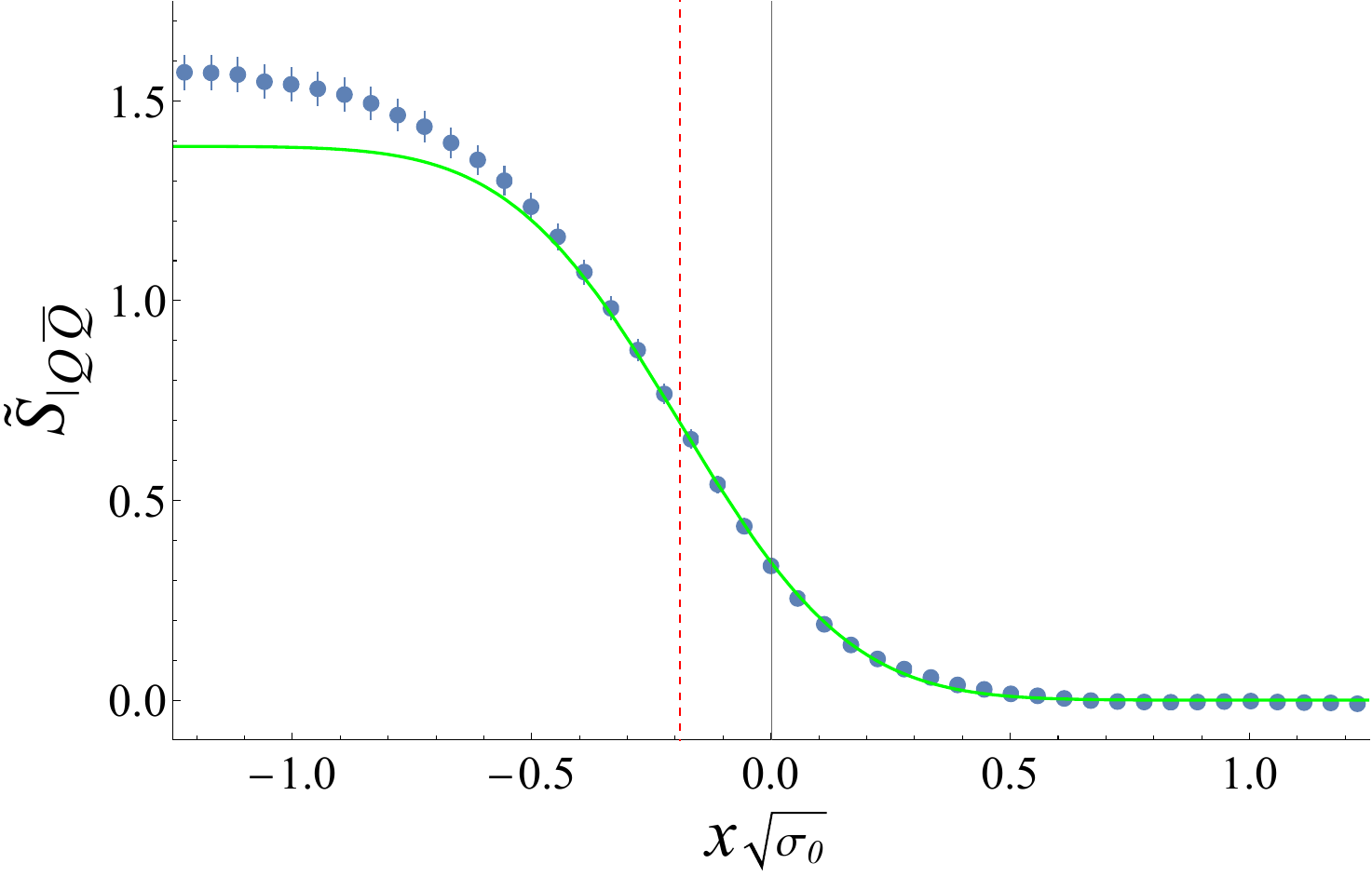}\\
\caption{\label{fig:refinedhalfslab}
Left:
A depiction of the ``half-slab'' geometry, with region $V$ (shaded blue) having width $w$ in the $y$ direction.
The quark and antiquark sources are separated by distance $L$ and are located at $(0,\pm L/2)$.
Right:
\FTEE in (2+1)D $SU(2)$ Yang-Mills theory and error function fit as a function of $x$ obtained by Ref. \cite{Amorosso:2024leg}.
The curve is centered at $x_c\sqrt{\sigma_0}=-0.19$ indicated by the red dashed line and has half-width $W\sqrt{\sigma_0}\approx 0.3$.
Data was obtained at $a\sqrt{\sigma_0}=0.056$ for $L=12a$, $w=4a$, and $y=0$.
Figures reproduced from \cite{Amorosso:2024leg}.
}
\end{figure}

The internal entanglement entropy of the flux tube is independent of the distance between the
quark and the antiquark in (1+1)D.
However in higher dimensional systems, the flux tube can fluctuate in transverse dimensions to avoid region $V$, or cross the boundary multiple times.
In this case,  $S_{\rm internal}$ takes form \cite{Amorosso:2024leg}
\begin{equation}
\label{eq:intersectionProbability}
S_\text{internal}=\langle F\rangle \cdot \log(N_c)\,,
\end{equation}
where $\langle F\rangle$ is the average number of times the flux tube crosses the $V/\bar V$ spatial boundary.
In Ref. \cite{Amorosso:2024leg}, the half slab geometry in Fig. \ref{fig:refinedhalfslab} is used to investigate \FTEEformat, where region $V$ consists of a finite width slab extending indefinitely to $x\to\infty$ (on a finite periodic lattice, the slab spans $L_x/2$ of the lattice).
As small deflections of the flux tube follow a roughly Gaussian distribution,
we anticipated $\langle F \rangle$ to be described by the error function of $x$ in this half slab geometry.

\FTEE indeed behaves like an error function, but offset from the center by a distance that was suggested \cite{Amorosso:2024leg} might be associated with the intrinsic width of the flux tube.
For a thin (zero intrinsic width) string, the number of boundary crossings $F$ is unambiguous, but for a flux tube with finite non-zero intrinsic width, it is unclear how to treat partial boundary crossings a priori. 
When only part of the flux tube intersects region $V$ (a ``partial boundary crossing''), this could in principle increment $F$ by zero, one, or some fraction proportional to how much of the flux tube enters $V$.
The results of Ref.~\cite{Amorosso:2024leg} seem to suggest that the flux tube
contributes to the entanglement entropy of a region only when its entire intrinsic width lies in that region.
We will investigate this further in the following section.

\section{2+1 D Yang-Mills simulations of FTE$^2$
  \label{sec:results}}
\begin{figure}[t!]
\centering
\includegraphics[width=.45\textwidth]{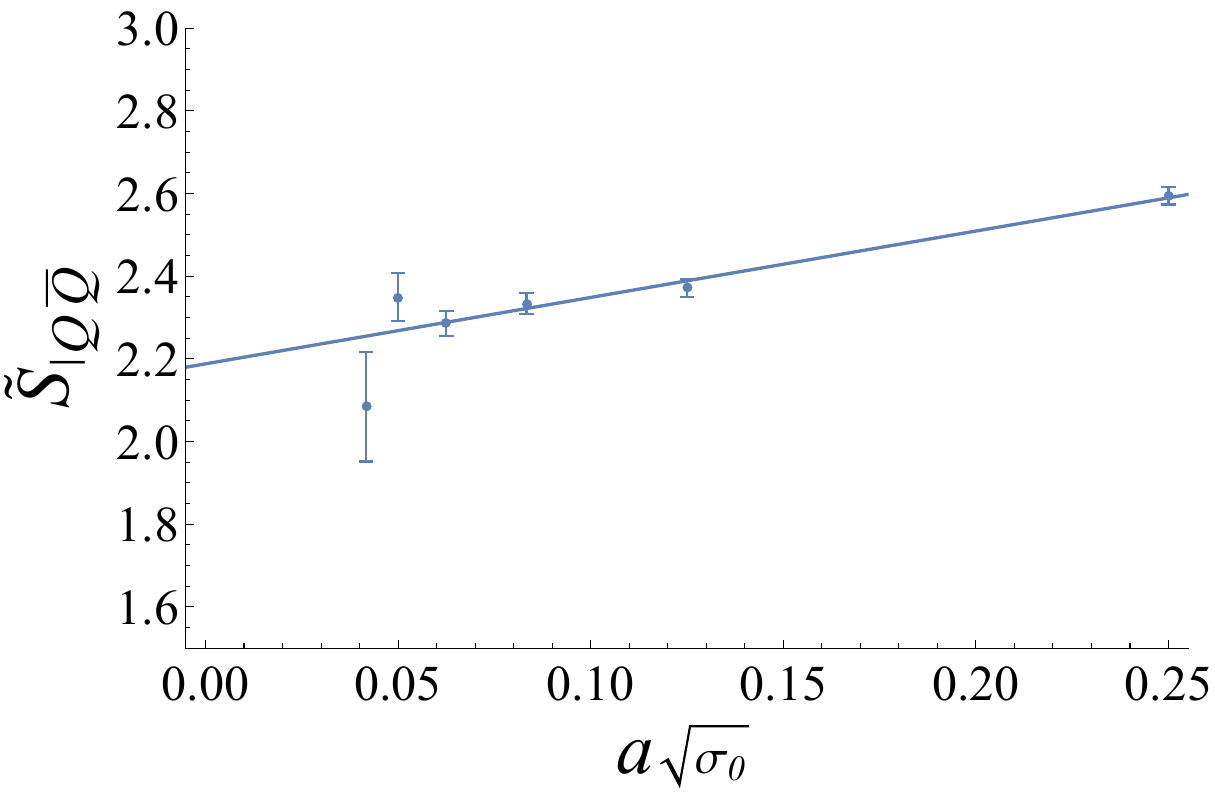}
\hfill
\includegraphics[width=.45\textwidth]{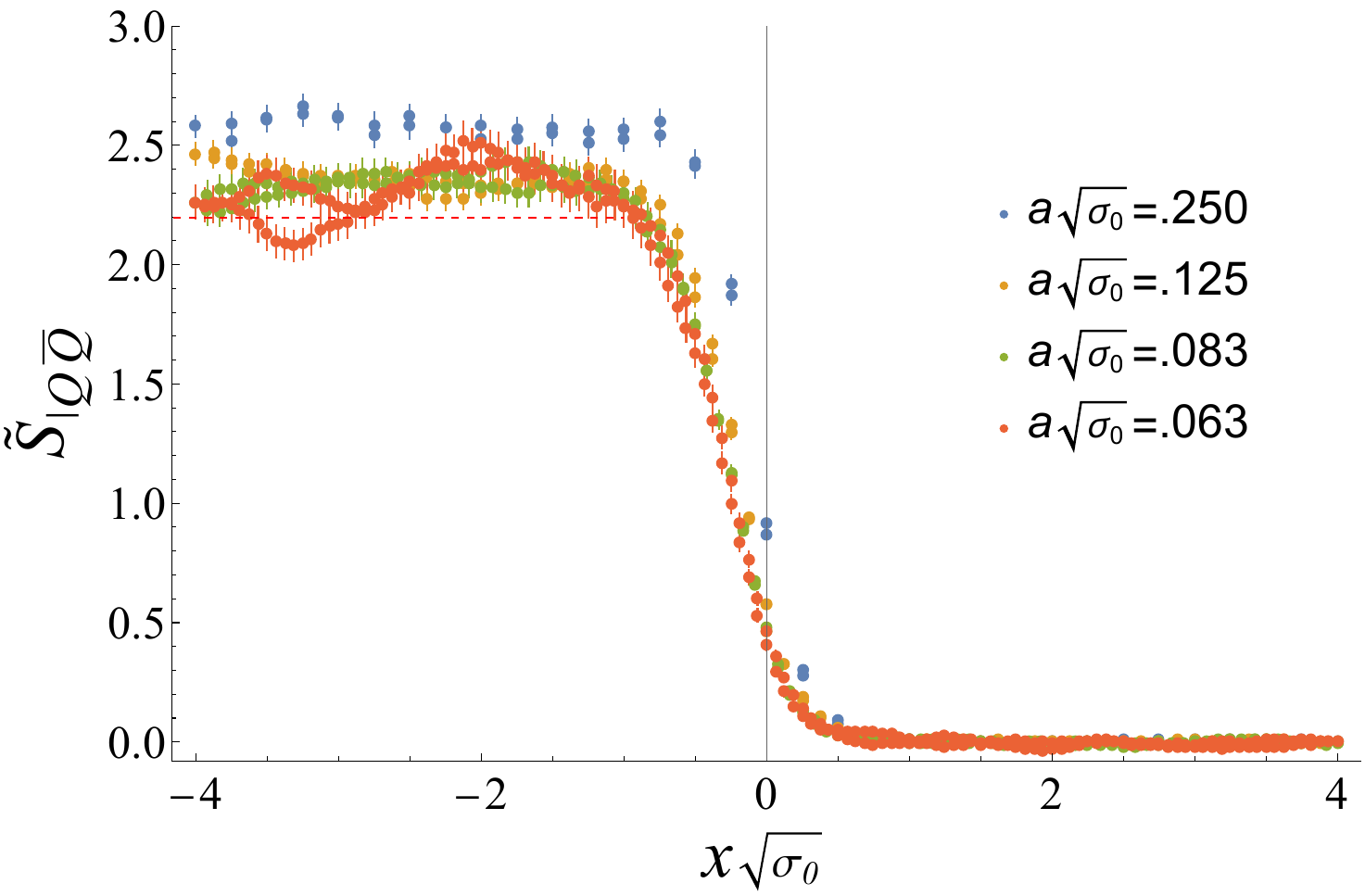}
\\
\caption{\label{fig:su3entropyascaling} 
Left: \FTEE $\tilde{S}_{\vert\QQbar}$ in (2+1)D $SU(3)$ Yang-Mills theory as a function of $a$ in the $x\to-\infty$ limit ($x\sqrt{\sigma_0}<-1.8$). Here $x$ is the relative position of the flux tube and region $V$, corresponding to the half-slab geometry shown in Fig. \ref{fig:refinedhalfslab}. Results are shown for a fixed $\QQbar$ distance $L\sqrt{\sigma_0}\approx0.75$, slab width $w\sqrt{\sigma_0}\approx0.25$, and $y=0$.
Right: FTE$^2$ as a function of $x$ in the half slab geometry. Other parameters are the same as in the figure on the left. The dashed red line is a constant $2\log3$, the expected internal entropy contribution in the $x\to -\infty$ limit. 
}
\end{figure}

We will present here preliminary calculations of \FTEE in $SU(2)$ and $SU(3)$ Yang-Mills gauge theory on a (2+1)D lattice, comparing \FTEE to the expectations from the internal entropy~(\ref{eq:intersectionProbability}) by varying the number of colors and boundary crossings.
We use the standard Wilson plaquette Yang-Mills action, performing sweeps with alternating
Kennedy-Pendleton heatbath updates \cite{Kennedy:1985nu} and over-relaxation updates
\cite{Brown:1987rra} and using the multilevel algorithm~\cite{Luscher:2001up} 
for computing the Polyakov loop correlators.
$N_c=2$ simulations are performed at $T_c/2$ to match the analysis in \cite{Amorosso:2024leg}.
$N_c=3$ simulations are performed at $T_c/4$.
%%%%%%%%%%%%%%%%%%%%%%%%%%%%%%%%%%%%%%%%%%%%%%%%%
We begin by studying $N_c=3$.
The resulting scaling study is shown in Fig.~\ref{fig:su3entropyascaling}. 
We observe similar qualitative behavior for $N_c=3$ as the $N_c=2$ case studied in Ref.~\cite{Amorosso:2024leg}, with the value of \FTEE in the $x\to -\infty$ limit once again indicating that it must be dominated by the internal color contribution $2\log N_c\approx2.20$ for string length $L\sqrt{\sigma_0}=0.75$.
We also observe that \FTEE at $x\gtrsim0$ takes approximately the form of the error function,
as anticipated from our discussion of Eq.~(\ref{eq:intersectionProbability}).
\FTEE is once again not symmetric around $x=0$. 
Fitting \FTEE data for $x\sqrt{\sigma_0}>-0.4$ to an error function $\sim (2\log N_c) \erf\big((x_c-x)/(W\sqrt{2})\big)$, we find its center at $x_c\sqrt{\sigma_0}=-0.27$ and its Gaussian half-width $W\sqrt{\sigma_0}\approx0.3$.
The half-width $W$ of the profile with $N_c=3$ matches its counterpart with $N_c=2$, 
while the displacement $x_c$ of the profile center is suggestively $\sim3/2$ of the case with $N_c=2$.
\begin{figure}[t!]
\centering
\includegraphics[width=.4\textwidth]{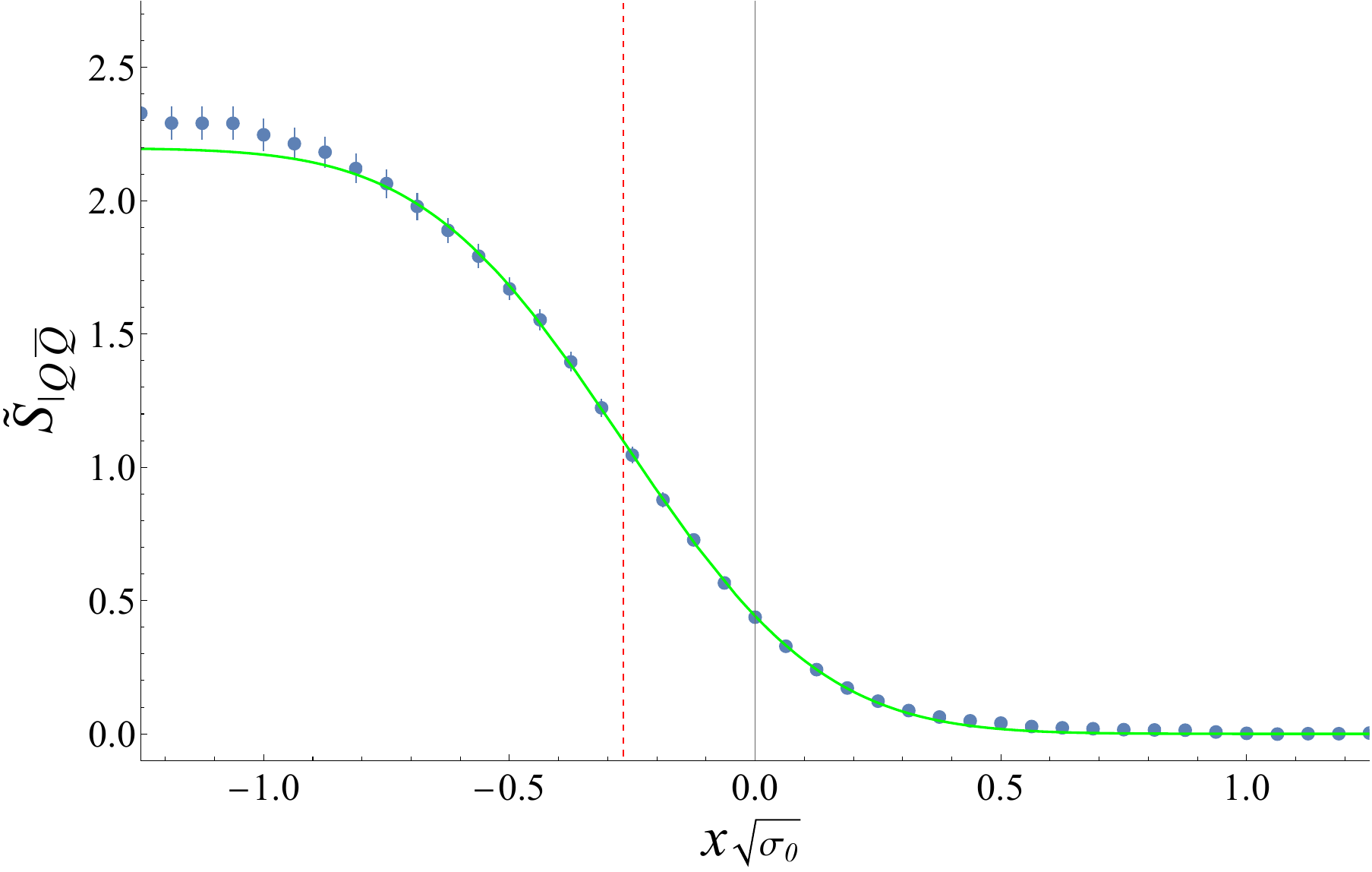}\\
\caption{\label{fig:errorFunctionComp3}
\FTEE in (2+1)D $SU(3)$ Yang-Mills theory as a function of $x$ obtained at $a\sqrt{\sigma_0}=0.063$ for $L=12a$,  $w=4a$, and $y=0$.
The green solid line is obtained by fitting the $x\sqrt{\sigma_0}>-0.4$ data to an error function normalized to $2\log N_c$ ($N_c=3$) 
by varying the offset $x_c$ and the half-width $W$.
The optimal error function curve is centered at $x_c\sqrt{\sigma_0}=-0.27$ indicated by the red dashed line and has half-width $W\sqrt{\sigma_0}\approx 0.3$.
}
\end{figure}
%---------------
%---------------------
%------------------------------------------------

\begin{figure}[t!]
\centering
\includegraphics[width=.45\textwidth]{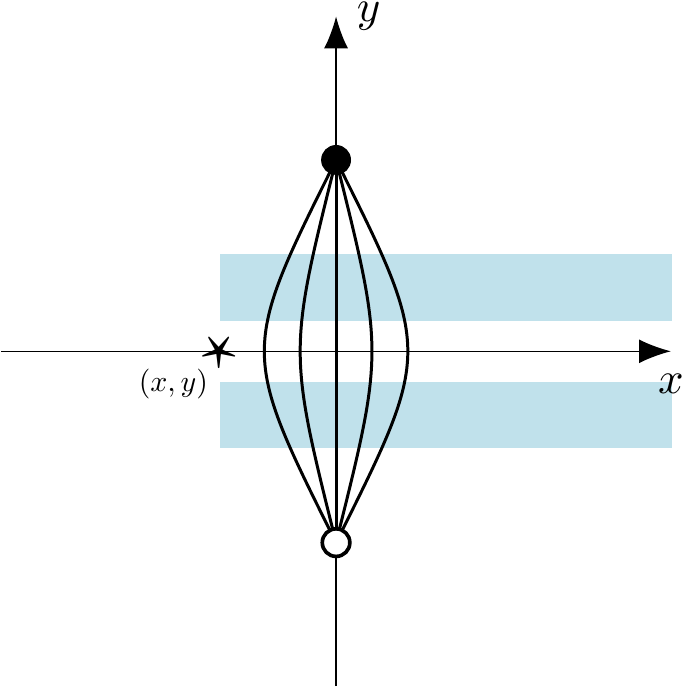}
\hfill
\includegraphics[width=.45\textwidth]{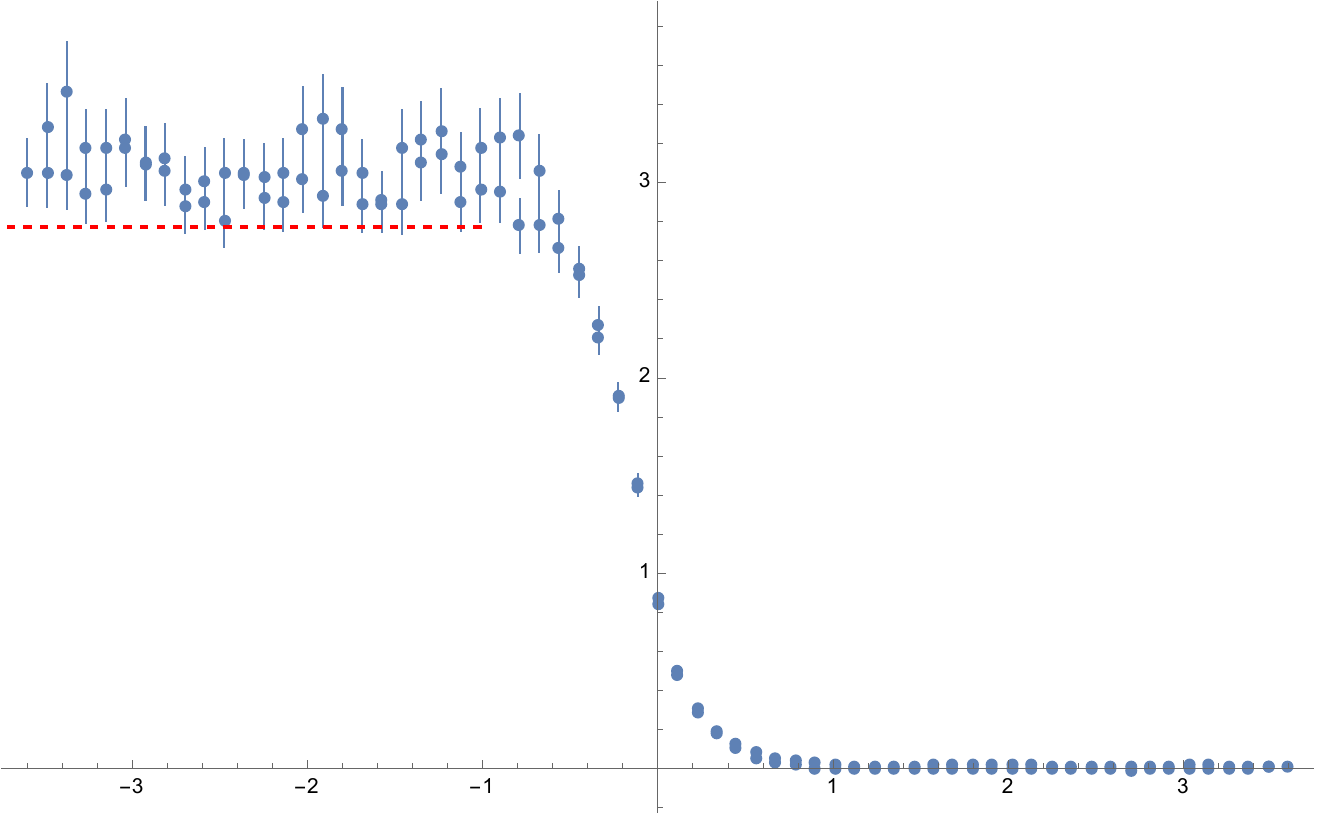}\\
\caption{\label{fig:refineddoubleslab}
Left:  ``Double-slab'' geometry, with region $V$ (shaded blue) consisting of two disconnected slabs of width $w$, separated by distance $w_{\rm gap}$.
Right:
\FTEE as a function of $x$ obtained for the double slab geometry at $a\sqrt{\sigma_0}=0.112$ for $L=10a$,  $w=w_{\rm gap}=2a$, and $y=0$.
The red dashed line is at $4\log N_c$ ($N_c=2$), the expectation of the internal entropy
contribution in the $x\to-\infty$ limit.
}
\end{figure}

Next, we examine the \FTEE dependence on the number of boundary crossings of the flux tube. 
Instead of one, we instead put two half-slabs in between the quark and the antiquark, doubling the number of boundary crossings in the $x\to-\infty$ limit for $N_c=2$.
The results of this study are shown in Fig.~\ref{fig:refineddoubleslab}, where the additional boundary crossings increase \FTEE in this limit to roughly its internal entropy prediction, $4\log(N_c)=4\log(2)$.
Our results generally agree with the hypothesis that the \FTEE consists of a dominant internal term taking the form in Eq.~(\ref{eq:intersectionProbability}) and a smaller vibrational component.

\begin{figure}[ht!]
\centering
\raisebox{-.5\height}{\includegraphics[width=.23\textwidth]{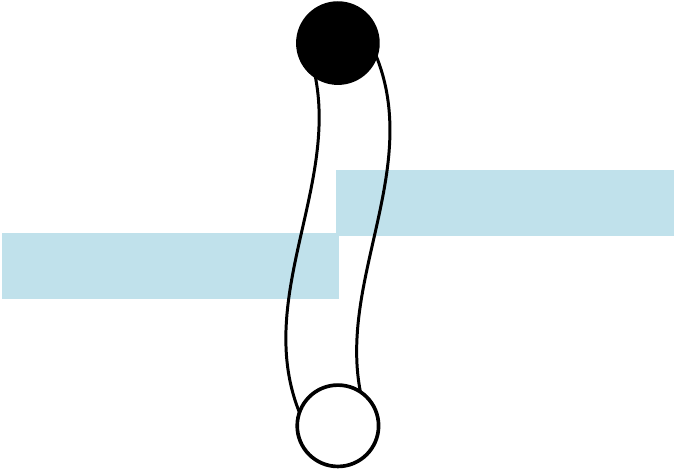}}
\hfill
\raisebox{-.5\height}{\includegraphics[width=.23\textwidth]{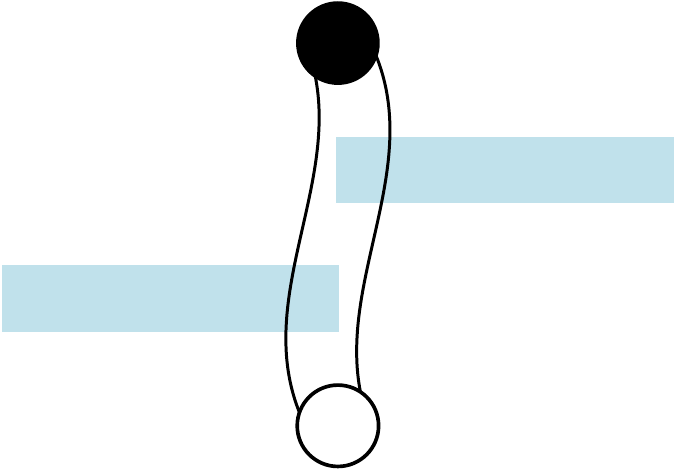}}
\hfill
\raisebox{-.5\height}{\includegraphics[width=.44\textwidth]{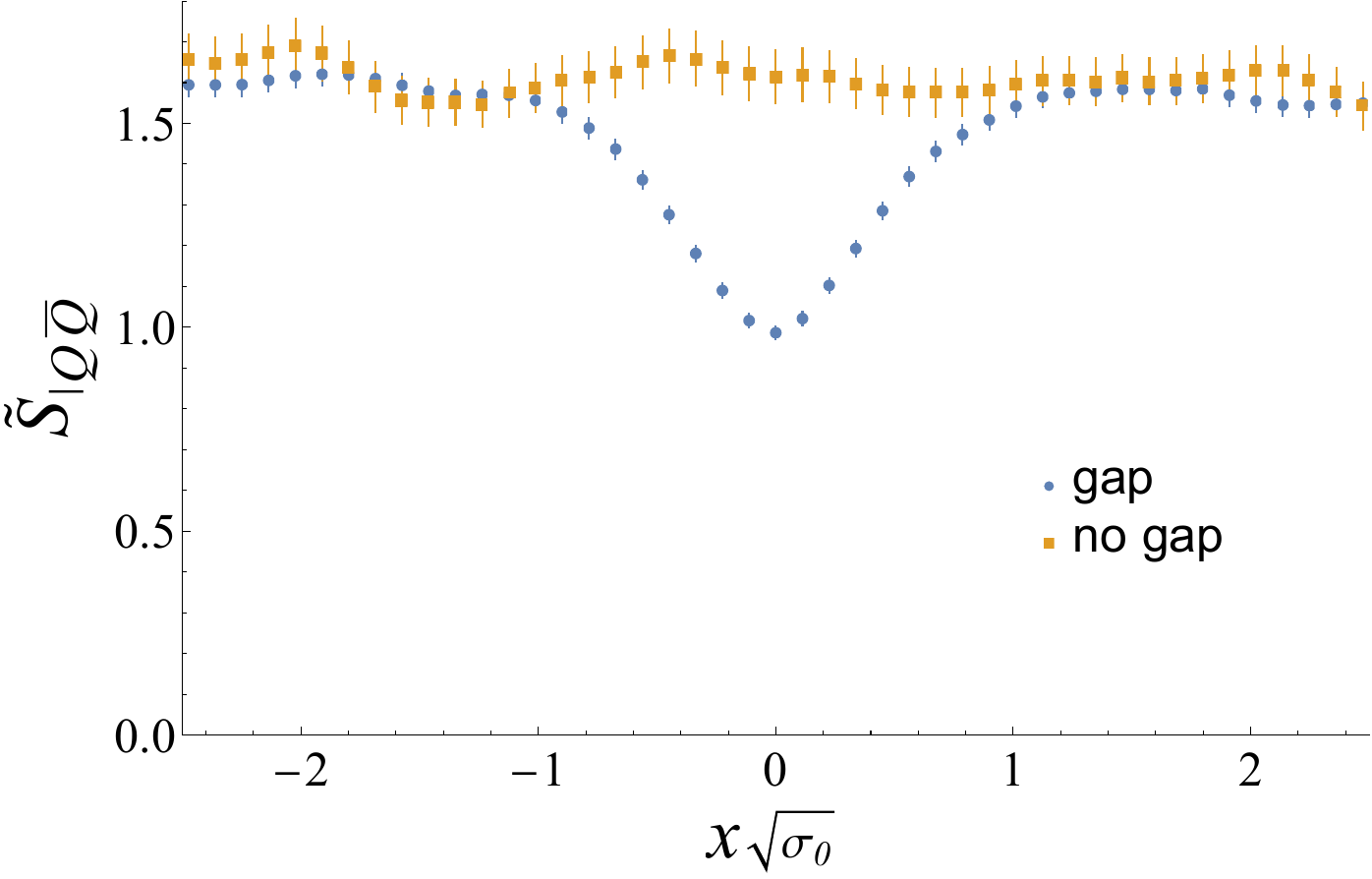}}
\caption{\label{fig:gapnogapData}
(Left) Schematic of a flux tube exhibiting a full boundary crossing in the ``no gap'' geometry. Here the flux tube is divided into two disconnected components in $\bar{V}$.
(Middle) Schematic of a flux tube exhibiting a partial boundary crossing in the ``gap'' geometry. Here the flux tube has one connected component in $\bar{V}$.
(Right) \FTEE of the staggered slab geometries with and without a gap. Data is obtained with $a\sqrt{\sigma_0}=0.112$, $y=0$, $L=10a$, and $N_c=2$. Each slab has width $2a$ and the gap has width $2a$.
}
\end{figure}

We investigated further the $x$-axis offset of the internal entropy.
In \cite{Amorosso:2024leg}, we suggested that the internal color degrees of freedom 
contribute to \FTEE only when the entire flux tube, including its finite intrinsic width, crosses into the complement region. 
In other words, only those configurations, in which the flux tube is fully intersected by the $V/\bar V$ boundary and divided by it into completely disconnected longitudinal segments, contribute to its internal entanglement.
This can be investigated by comparing the two geometries shown in Fig.~\ref{fig:gapnogapData} (left, middle), where region $V$ consists of two half-slabs extending in opposite $x$ directions.
Since \FTEE has very little $y$-dependence in the single-slab case \cite{Amorosso:2024leg}, naively one might expect the two slab geometries in Fig.~\ref{fig:gapnogapData} to produce similar \FTEE values.
Indeed, if the flux tube had infinitesimal intrinsic width, configurations with $F=0$ (no crossings) and $F=4$ would have had \emph{equal probabilities} in the gapped geometry (Fig.~\ref{fig:gapnogapData}, middle), which would result in the same \FTEE value as the ``no-gap'' geometry (Fig.~\ref{fig:gapnogapData}, left).
However, if the flux tube has finite width and the above hypothesis is true, we expect substantially different results from these two geometries:
the ``staggered slab'' geometry without a gap always partitions the flux tube into two or more disconnected components in $\bar{V}$, while the ``staggered slab'' geometry with a gap allows for configurations of the flux tube at $x\sim0$ with one connected component in $\bar{V}$, as shown in Fig.~\ref{fig:gapnogapData} (middle).
The results of this study are shown in Fig~\ref{fig:gapnogapData} (right).
There is indeed a significant difference between the staggered slab geometry, with and without a gap, suggesting that partial boundary crossings do not contribute significantly to $F$ in Eq.~(\ref{eq:intersectionProbability}), thereby lowering the internal entropy in the ``staggered slab-with gap'' geometry.
This is indirect evidence that \FTEE can be used to study the intrinsic width of a color flux tube.

%%%%%%%%%%%%%%%%%%%%%%%%%%%%%%%%%%%%%%%%%%%%%%%%%%%%%%%%%%%%%%%%%%%%%%%%%%%%%%
\section{Discussion}

In this work, we continued our study of flux tube entanglement entropy, investigating its dominant component, the entanglement  due to internal colorful degrees of freedom of the flux tube.
We further tested the conjectured form  of the internal entropy in Eq.~(\ref{eq:intersectionProbability}), demonstrating that it successfully describes \FTEE for $N_c=3$, and number of boundary crossings $F=4$,  in addition to the previously studied $N_c=2$ and $F=2$ cases.
We further evaluated the specific dynamics of a flux tube partially intersecting region $V$, finding a significant difference between flux tubes that fully and partially intersect region $V$, due to the intrinsic width of the flux tube.
Our results suggest that it is important that effective string models of confinement in pure gauge theories take into account the intrinsic width of the QCD string, in line with prior work in \cite{Caselle:2012rp,Verzichelli:2025cqc}. These preliminary conclusions can be further tested by extending our study to more nontrivial geometries, and to Yang-Mills theory in 3+1 dimensions.

%%%%%%%%%%%%%%%%%%%%%%%%%%%%%%%%%%%%%%%%%%%%%%%%%%%%%%%%%%%%%%%%%%%%%%%%%%%%%%
\section*{Acknowledgements}
R.A is supported by the Simons Foundation under Award number 994318 (Simons Collaboration on Confinement and QCD
Strings). S.S. is supported by NSF supported by the National Science Foundation under award PHY-2412963. 
In addition, R.A. is supported in part by the Office of Science, Office of Nuclear Physics,
U.S. Department of Energy under Contract No. DEFG88ER41450 and by the National Science Foundation under award
PHY-2412963.
R.V is supported by the U.S. Department of Energy, Office of Science under contract DE-SC0012704. 
R.V's work on quantum information science is supported by the U.S. Department of Energy, Office of Science, National
Quantum Information Science Research Centers, Co-design Center for Quantum Advantage (C$^2$QA) under contract number
DE-SC0012704. 
R.V. was also supported at Stony Brook by the Simons Foundation as a co-PI under Award number 994318 (Simons Collaboration on Confinement and QCD Strings). 
The authors thank Stony Brook Research Computing and Cyberinfrastructure and the Institute for Advanced Computational
Science at Stony Brook University for access to the Seawulf HPC system, which was made possible by grants from the
National Science Foundation (awards 1531492 and 2215987) and matching funds from the Empire State Development’s Division
of Science, Technology and Innovation (NYSTAR) program (contract C210148).

\bibliography{entent-lat23proc}
\bibliographystyle{JHEP}

\end{document}